\documentclass[fleqn,usenatbib]{mnras}

\usepackage{newtxtext,newtxmath}

\usepackage[T1]{fontenc}

\DeclareRobustCommand{\VAN}[3]{#2}
\let\VANthebibliography\thebibliography
\def\thebibliography{\DeclareRobustCommand{\VAN}[3]{##3}\VANthebibliography}

\usepackage{graphicx}	
\usepackage{breakurl}

\newcommand{\kms}{~km s$^{-1}$~}

\newcommand{\WR}{WR~48a~}
\newcommand{\WRE}{WR~48a}
\newcommand{\dotMyr}{~M$_{\odot}$~yr$^{-1}$~}
\newcommand{\XMM}{{\it XMM-Newton~}}
\newcommand{\XMME}{{\it XMM-Newton}}
\newcommand{\Chandra}{{\it Chandra~}}
\newcommand{\ChandraE}{{\it Chandra}}
\newcommand{\Swift}{{\it Swift~}}

\newcommand{\xspec}{{\sc xspec~}}
\newcommand{\xspecE}{{\sc xspec}}

\title[CSWs in \WRE]
{ \Chandra revisits \WRE: testing colliding wind models in massive
binaries
}

\author[S.A.Zhekov, M.Gagn\'{e} \& S.L.Skinner]
{Svetozar A. Zhekov$^1$\thanks{ E-mail: szhekov@astro.bas.bg}, 
Marc Gagn\'{e}$^2$ and Stephen L. Skinner$^3$\\
$^1$Institute of Astronomy and National Astronomical Observatory
(Bulgarian Academy of Sciences),\\
72 Tsarigradsko Chaussee Blvd., Sofia 1784, Bulgaria\\
$^2$Department of Earth and Space Sciences, West Chester University,
West Chester, PA 19383, USA\\
$^3$Center for Astrophysics and Space Astronomy (CASA), 
University of Colorado, Boulder, CO 80309-0389, USA\\
}

\date{}

\pubyear{2020}

\begin{document}
\label{firstpage}
\pagerange{\pageref{firstpage}--\pageref{lastpage}}
\maketitle

\begin{abstract}
We present results of new \Chandra High-Energy Transmission Grating
(HETG) observations (2019 November - December) of the massive 
Wolf-Rayet binary \WRE. 
Analysis of these high-quality data showed that the spectral lines in 
this massive binary are broadened (FWHM = 1400\kms) and 
marginally blueshifted  ($\sim -100$\kms).
A direct modelling of these high-resolution spectra in
the framework of the standard colliding stellar wind (CSW) picture 
provided a very good correspondence between the shape of the 
theoretical and observed spectra. Also, the theoretical 
line profiles are in most cases an acceptable representation of 
the observed ones.
We applied the CSW model to the X-ray spectra of \WR from previous 
observations: \ChandraE-HETG (2012 October) and \XMM (2008 January). 
From this expanded analysis, we find that
the observed X-ray emission from \WR is variable on the long timescale
(years) and the same is valid for its intrinsic X-ray emission. This
requires variable mass-loss rates over the binary orbital period. 
The X-ray absorption (in excess of that from the stellar winds 
in the binary) is variable as well.
We note that lower
intrinsic X-ray emission is accompanied by higher X-ray
absorption.
A qualitative explanation could be that the presence of clumpy and
non-spherically symmetric stellar winds may play a role.
\end{abstract}




\section{Introduction}
\WR is one of the five objects originally classified as episodic dust 
makers amongst the carbon-rich (WC) Wolf-Rayet (WR) stars in the 
Galaxy \citep{williams_95}.
As reported by \citet{danks_83} and \citet{danks_84}, 
this WC star was discovered in a near-infrared survey of the
Sagitarius arm of the Galaxy. Given its proximity (within $2'$) to the
compact clusters Danks 1 and 2 in the  G305 star-forming
region, \WR is likely a member of one of these clusters.

It is currently assumed that the episodic (especially periodic)
dust formation in WC stars is result of colliding stellar winds
(CSW) near periastron passage in wide WR binaries whose orbits have
high eccentricity. This is particularly illustrated by the properties
of WR 140, considered the prototype object of CSW binaries (e.g.,
\citealt{williams_90}; \citealt{williams_08}).
Also, it is worth recalling that CSW binaries are expected to have
enhanced X-ray emission, as originally proposed by \citet{cherep_76}
and \citet{pri_us_76} and as illustrated by the first systematic X-ray
survey of WRs with the {\it Einstein} Observatory (\citealt{po_87}; 
for a review on the progress of observational and theoretical studies 
of X-ray emission from CSW massive binaries of early spectral types,
see \citealt{rauw_naze_16}).

We note that the study by \citet{williams_12} revealed a recurrent 
dust formation in \WR on a timescale of more than 32 years. Such a
finding indicates that \WR is very likely a wide CSW binary.
On the other hand, 
analysis of the \XMM and \Chandra spectra of \WR provided additional
support to the CSW picture in this object (\citealt{zhgsk_11};
\citealt{zhgsk_14}).
Namely, it was found that
the X-ray emission of \WR is of thermal origin.
It is variable on a long timescale (years) and 
the same is valid for the X-ray absorption to this object.
This is the most X-ray luminous WR star in the Galaxy detected so
far (L$_X\sim$ 10$^{35}$ ergs s$^{-1}$), after the black-hole 
candidate Cyg X-3.
Recently, \WR is classified as WC8 + WN8 massive binary\footnote{Galactic Wolf
Rayet Catalogue;
\url{http://pacrowther.staff.shef.ac.uk/WRcat/index.php}}.
It is worth recalling that the single 
WC stars are very faint or X-ray quiet objects 
(\citealt{os_03}; \citealt{sk_06}), and single WN8 stars are 
faint in X-rays (L$_X <$ 10$^{32}$ ergs s$^{-1}$; 
\citealt{gosset_05}; \citealt{sk_12}; \citealt{sk_21}).
Thus, binarity should play a key role for the X-ray emission from
\WRE.

So, to address the physical picture of CSWs in \WR in some 
detail, we need to confront the observational data with the 
corresponding theoretical model predictions. For this reason, we
planned new X-ray observations of \WR with high spectral resolution
that were expected to provide X-ray spectra with better quality than
achieved in the earlier observations of this object. 

This work
provides the results from the first direct modelling of the 
X-ray emission from \WR in the framework of the CSW picture.

\section{Observations and data reduction}
\label{sec:data}
\Chandra revisited \WR in three occasion during 2019 November-December
with a total effective exposure of 94.4 ks: \Chandra Obs ID 21162
(November 27; 28.63 ks), 23085 (November 28; 28.61 ks), 22938
(December 26; 37.16 ks). The observations were carried out with the 
high-energy transmission gratings (HETG). 
We extracted the corresponding first-order and zero-order X-ray 
spectra of \WR as recommended in the Science Threads for Grating
Spectroscopy in the
{\sc ciao} 4.12\footnote{Chandra Interactive Analysis of Observations
(CIAO), \url{https://cxc.harvard.edu/ciao/}.} data analysis software
and using the \Chandra calibration database {\sc calbd} v.4.9.4.

We closely inspected the zeroth order images and no second source was
present in vicinity of \WR as claimed to be found in the heavily piled
up observation of \WR (\Chandra Obs ID 8922; 2008 December 13): 
for details on the presumed second source,  see section 4.11 and 
fig. 31 in \citet{townsley_19}, and it should be also noted that no
second source was present in the \Chandra observation of 2012 October
(\Chandra Obs ID 13636).

In this study, we focus on the 
first-order Medium Energy Grating (MEG) and High Energy Grating (HEG) 
spectra.
Since no appreciable variability was detected 
(the differences of count
rates between every two data sets are within their corresponding
$1\sigma$ values),
we constructed total HETG
spectra of \WR with a total number of source counts of 
7738 (MEG), 4592 (HEG), 11030 (HETG-0, zeroth order).
We may refer to this data set as `Chandra 2019'  throughout the text.

The spectral analysis was performed using
standard as well as custom models in version 12.10.1 of
\xspec \citep{Arnaud96}.

\begin{figure*}
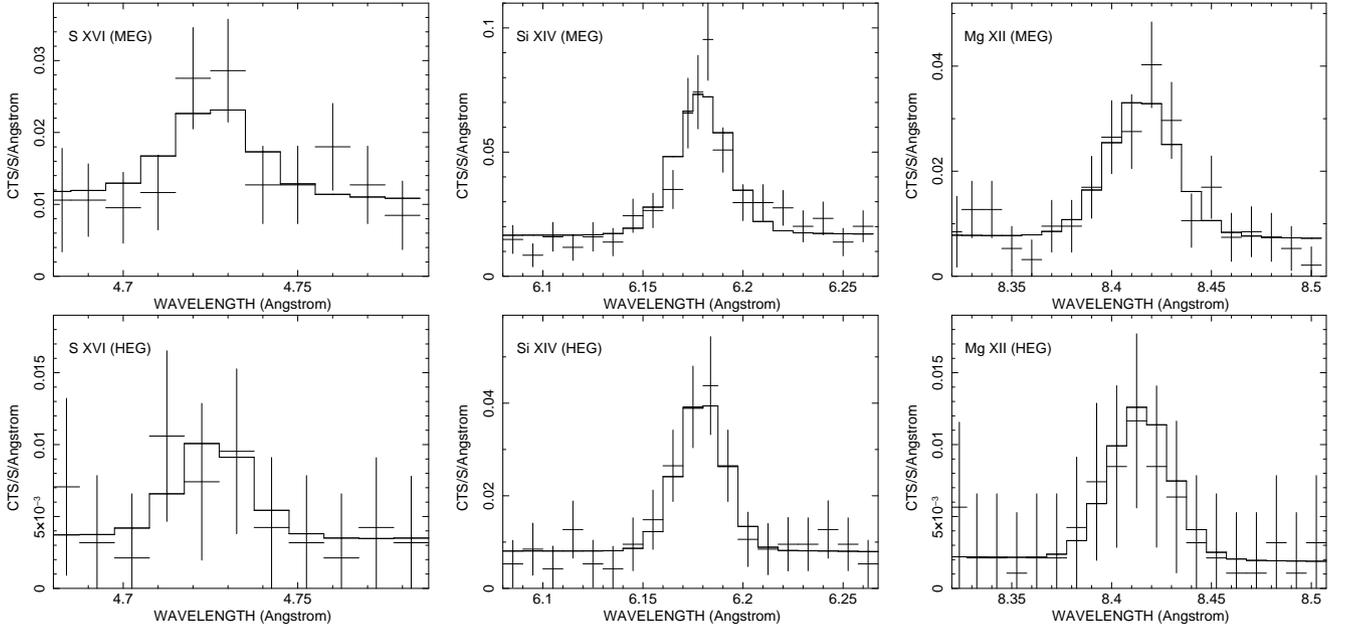

\begin{center}
\includegraphics[width=1.6in,height=2.3in,angle=-90]{fig1a.eps}
\includegraphics[width=1.6in,height=2.3in,angle=-90]{fig1b.eps}
\includegraphics[width=1.6in,height=2.3in,angle=-90]{fig1c.eps}
\includegraphics[width=1.6in,height=2.3in,angle=-90]{fig1d.eps}
\includegraphics[width=1.6in,height=2.3in,angle=-90]{fig1e.eps}
\includegraphics[width=1.6in,height=2.3in,angle=-90]{fig1f.eps}
\end{center}
\caption{Line profile fits to the H-like doublets in the first-order
HETG spectra of \WRE. The spectra were rebinned for presentation
purposes.
}
\label{fig:lines_h-like}
\end{figure*}

\begin{figure*}
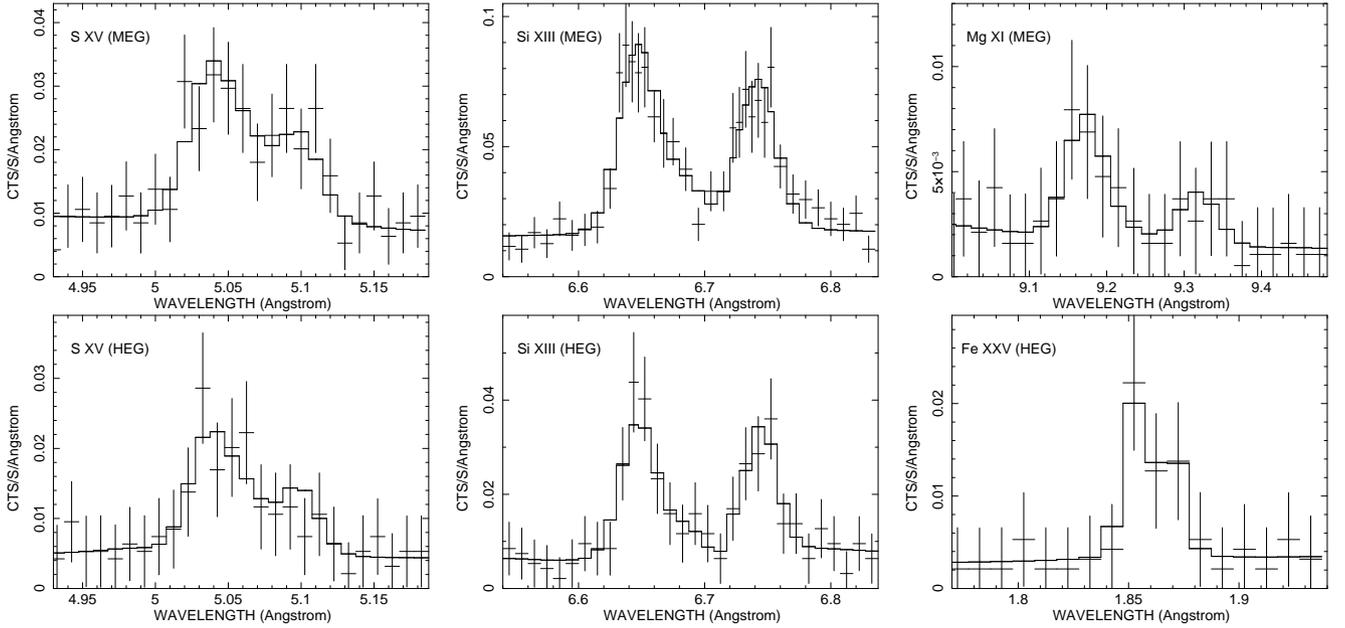

\begin{center}
\includegraphics[width=1.6in,height=2.3in,angle=-90]{fig2a.eps}
\includegraphics[width=1.6in,height=2.3in,angle=-90]{fig2b.eps}
\includegraphics[width=1.6in,height=2.3in,angle=-90]{fig2c.eps}
\includegraphics[width=1.6in,height=2.3in,angle=-90]{fig2d.eps}
\includegraphics[width=1.6in,height=2.3in,angle=-90]{fig2e.eps}
\includegraphics[width=1.6in,height=2.3in,angle=-90]{fig2f.eps}
\end{center}
\caption{The same as in Fig.~\ref{fig:lines_h-like} but for the
He-like triplets in the first-order HETG spectra of \WRE. The spectra
were rebinned for presentation
purposes.
}
\label{fig:lines_he-like}
\end{figure*}

\section{Spectral lines}
\label{sec:lines}
We note that the new Chandra 2019 observations 
of \WR provided high-resolution spectra with better quality 
compared to the previous such an observation (2012 October): the
number of source counts in the new MEG and HEG spectra is a factor of
3.1 - 3.7 higher, 
thanks to the \WR higher X-ray brightness in 2019.
So, we could analyse with acceptable accuracy more
spectral lines in order to deduce some kinematic information about the 
gas flows in this CSW binary.
To do so, we fitted the line profiles of the strong H-like doublets
of S XVI, Si XIV, Mg XII and He-like triplets of Fe XXV, S XV, 
Si XIII, Mg XI with the following model.

For the He-like triplets, the model was a sum of
three Gaussians and a constant continuum. The centres of the
triplet components were equal to their values given in
the  AtomDB data base (Atomic Data for
Astrophysicists)\footnote{For AtomDB, see http://www.atomdb.org/}.
All components had the same line width and line shift.
The free parameters of the fit were the common line shift, common line 
width, the individual fluxes of the three components and the continuum
level.
For the H-like doublets, we used a similar model but with a sum of two 
Gaussians and the component intensity ratios were fixed at their atomic 
data values.

\begin{figure*}
\begin{center}
\includegraphics[width=3.5in,height=2.5in]{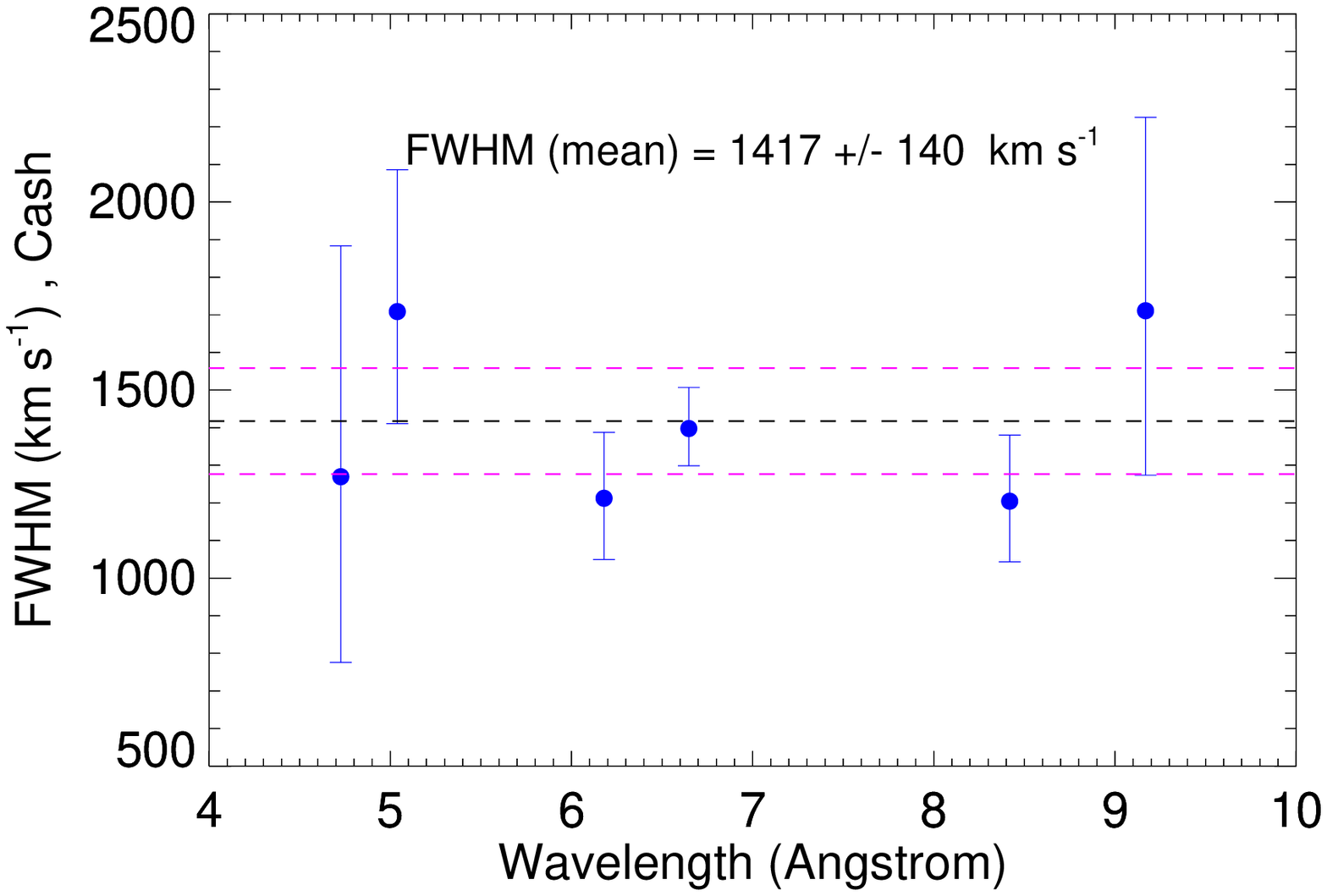}
\includegraphics[width=3.5in,height=2.5in]{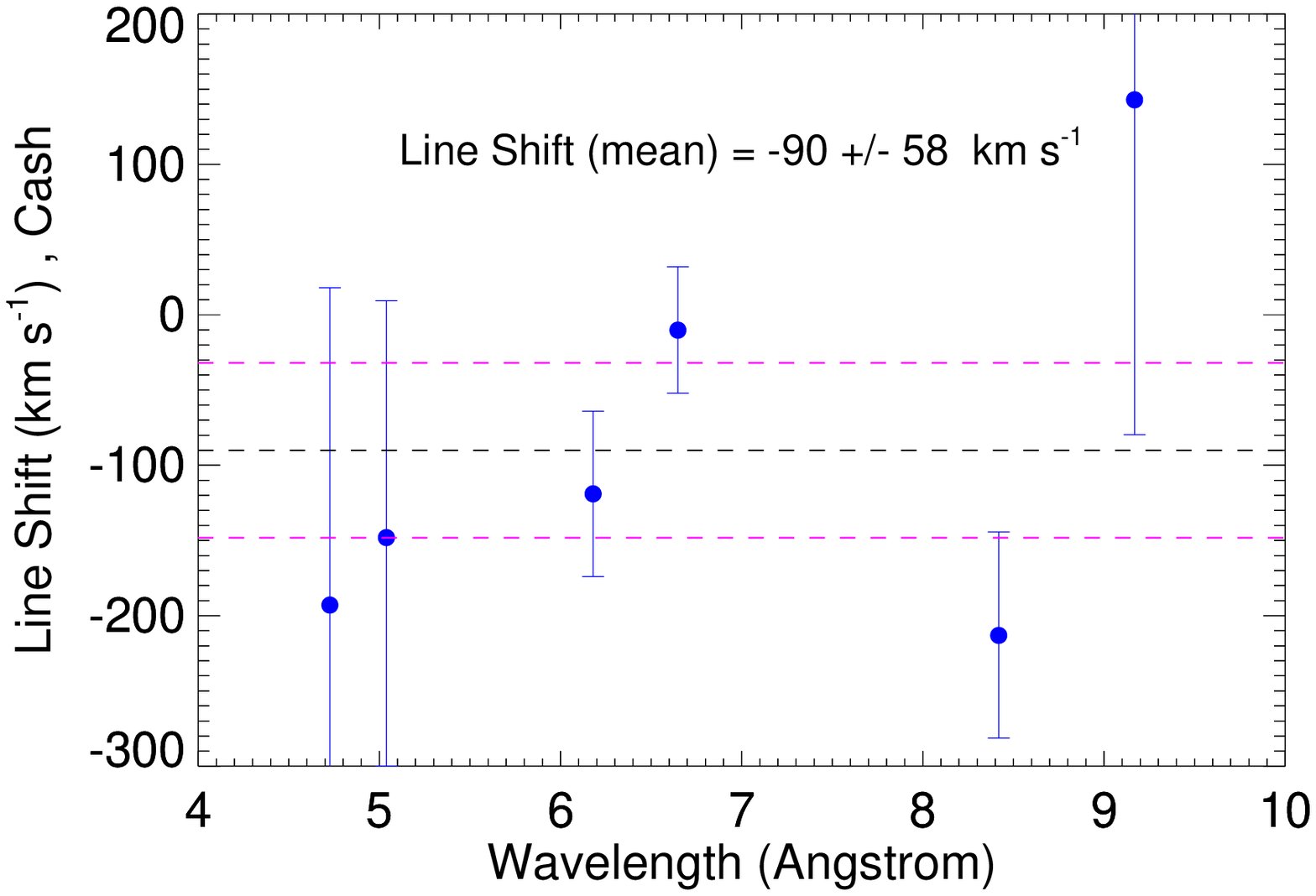}
\end{center}
\caption{Spectral line parameters \WRE.
}
\label{fig:lines_param}
\end{figure*}

\begin{table*}
\caption{Line Parameters
\label{tab:lines}}
\begin{tabular}{lccrrc}
\hline
\multicolumn{1}{l}{Line} & \multicolumn{1}{c}{$\lambda_{lab}^{a}$} &
\multicolumn{1}{c}{FWHM$^{b}$} &
\multicolumn{1}{c}{Line Shift$^{c}$} & \multicolumn{1}{c}{Flux$^{d}$}
& \multicolumn{1}{c}{Ratio}\\
\multicolumn{1}{c}{} & \multicolumn{1}{c}{(\AA)} &
\multicolumn{1}{c}{(\kms)} &
\multicolumn{1}{c}{(\kms)} & \multicolumn{1}{c}{}  &
\multicolumn{1}{c}{(ATOMDB)} \\
\hline
Fe XXV K$_{\alpha}$ & 1.850 &
0$^{e}$ & 398$^{e}$ & 25.78$^{+5.00}_{-4.58}$ \\
\,\,\,(i/r)  &  &  &  & 0.44$^{+0.37}_{-0.24}$ & 0.38 \\
\,\,\,(f/r)  &  &  &  & 0.60$^{+0.35}_{-0.24}$ & 0.30 \\
S XVI L$_{\alpha}$ & 4.727 &
1270$^{+614}_{-494}$ & -193$^{+211}_{-186}$ & 11.86$^{+3.82}_{-3.25}$
\\
S XV K$_{\alpha}$ & 5.039 &
1709$^{+377}_{-298}$ & -148$^{+157}_{-152}$ & 58.56$^{+5.53}_{-5.33}$
\\
\,\,\,(i/r) &  &  &  & 0.49$^{+0.25}_{-0.22}$  &  0.23 \\
\,\,\,(f/r)  &  &  &  & 0.64$^{+0.17}_{-0.14}$ &  0.44 \\
Si XIV L$_{\alpha}$ & 6.180 &
1212$^{+175}_{-163}$  & -119$^{+55}_{-55}$ &
25.93$^{+2.36}_{-2.26}$\\
Si XIII K$_{\alpha}$ & 6.648 &
1398$^{+109}_{-99}$ & -10$^{+42}_{-42}$ & 72.02$^{+3.58}_{-3.53}$ \\
\,\,\,(i/r)  &  &  &  & 0.27$^{+0.06}_{-0.05}$ &  0.20 \\
\,\,\,(f/r)  &  &  &  & 0.75$^{+0.07}_{-0.07}$ &  0.52 \\
Mg XII L$_{\alpha}$ & 8.419 &
1204$^{+176}_{-161}$  & -213$^{+69}_{-68}$ &
15.45$^{+1.87}_{-1.79}$\\
Mg XI K$_{\alpha}$ & 9.169 &
1711$^{+514}_{-437}$ & 143$^{+220}_{-223}$ & 16.11$^{+3.54}_{-3.36}$
\\
\,\,\,(i/r)  &  &  &  & 0.05$^{+0.23}_{-0.05}$ &  0.19 \\
\,\,\,(f/r)  &  &  &  & 0.46$^{+0.23}_{-0.17}$ &  0.59 \\
\hline

\end{tabular}

{\it Note}.
Results from the fits to the line profiles of emission lines in \WR
with the associated $1\sigma$ errors.
The first-order MEG and HEG spectra were fitted simultaneously for
S XVI, S XV, Si XIV, Si XIII, and  Mg XII lines, while only the HEG
and MEG data were used for Fe XXV and MG XI, respectively.
For the He-like triplets,
the flux ratios of the intercombination to the resonance line (i/r)
and of
the
forbidden to the resonance line (f/r) are given as well.
The Cash statistic \citep{cash_79} was adopted in the fits.
\\
$^{a}$ The laboratory wavelength of the main component.\\
$^{b}$ The line width (FWHM).\\
$^{c}$ The shift of the spectral line centroid.\\
$^{d}$ The observed total multiplet flux in units of $10^{-6}$
photons cm$^{-2}$ s$^{-1}$.\\
$^{e}$ Because of the poor photon statistics, these line parameters
are not
constrained.

\end{table*}

For each spectral line complex (H-like doublet or He-like triplet), we
fitted the MEG and HEG spectra simultaneously sharing the same model
parameters but the continuum level. In cases where the data quality 
was poor in either the MEG or HEG spectrum, we used only the better 
quality spectrum with more counts for fitting some He-like triplets:
Fe XXV (HEG) and Mg XI (MEG)

It is quite common that the high resolution X-ray spectra have a very
low number of counts (even zero counts) in the spectral bins.
The photon statistics could be improved by re-binning the X-ray
spectrum of a given object but this is at the expense of deteriorating
the spectral resolution. To avoid this, we worked with the unbinned
spectra (with no background subtracted)  and made use of the Cash 
statistic \citep{cash_79} as implemented in \xspecE.

Figures \ref{fig:lines_h-like}, \ref{fig:lines_he-like},
\ref{fig:lines_param} and Table \ref{tab:lines} present the results from
the fits to the line profiles in the grating spectra of \WR in 2019.
We note that the parameters of the Fe XXV He-like triplet are not
constrained with exception to the total observed line flux. On
average, the X-ray emission lines are marginally blueshifted by $\sim
100$\kms. 
On the
other hand, all the lines show a consistent line broadening of $\sim
1400$\kms.
Forbidden lines in the He-like triplet do not seem to be suppressed,
which is a sign that these spectral features form in hot plasmas with
relatively low density and located far from strong sources of UV
emission.  The latter could be considered as a possible indication
that these lines form in CSWs in wide massive binaries.

\begin{figure*}
\begin{center}
\includegraphics[width=\columnwidth]{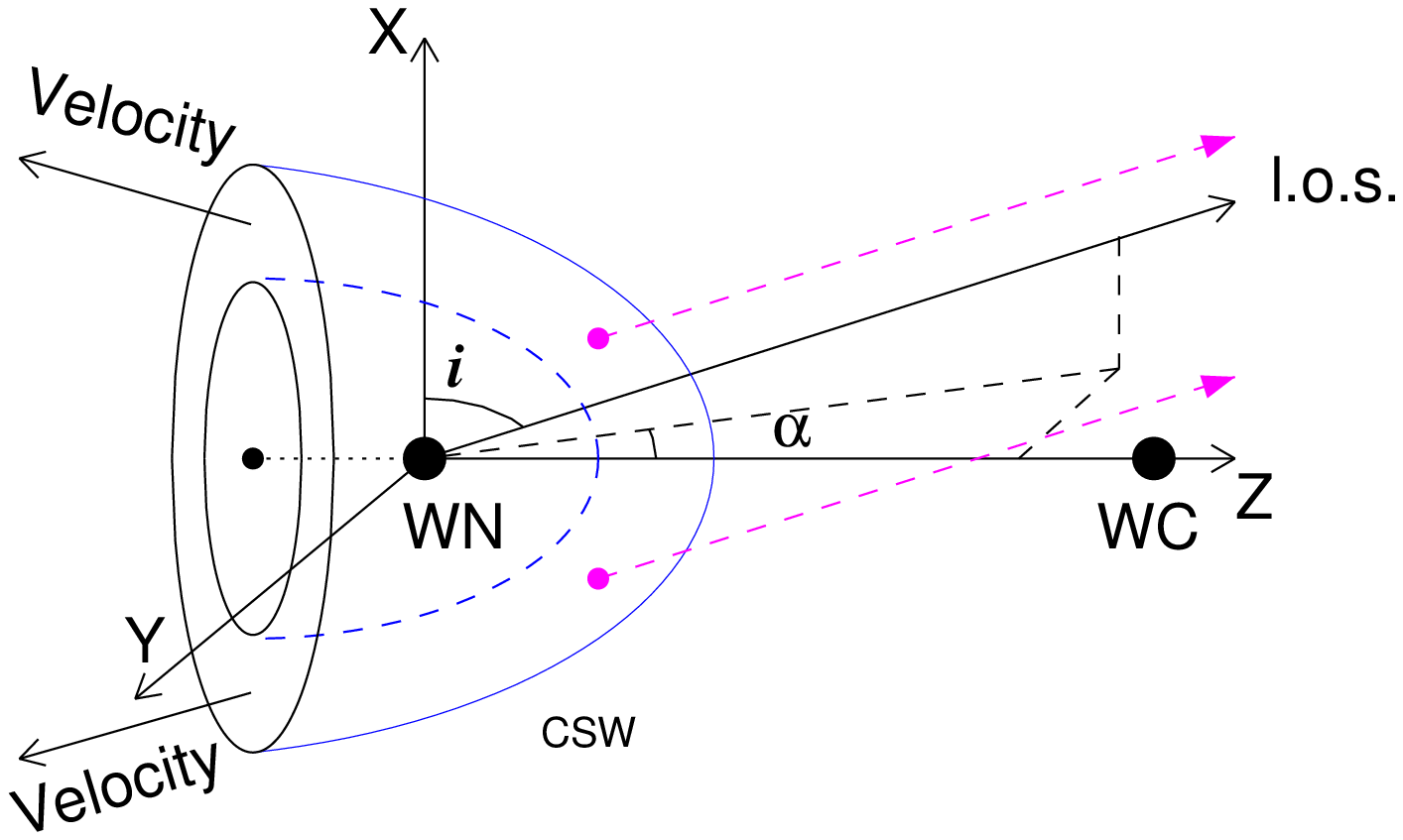}
\includegraphics[width=\columnwidth]{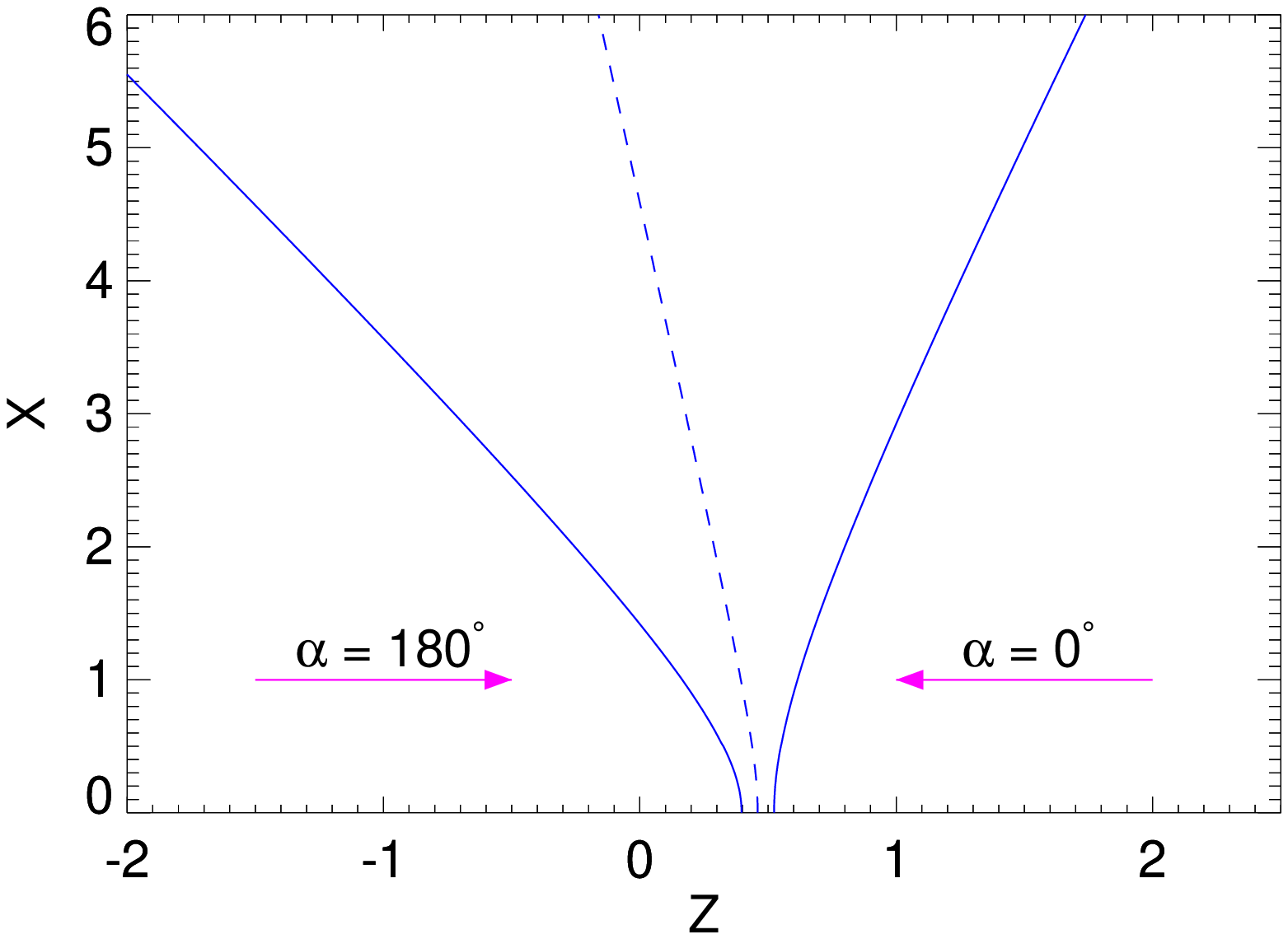}
\end{center}
\caption{
{\it
Left-hand panel:
}
schematic presentation of the CSW region in the WC$+$WN binary \WRE.
The label CSW marks the wind interaction region, which is a 3D
structure with its axis of symmetry Z. The axes X and Y complete the 
Cartesian coordinate system.
The two angles $i$ (orbital inclination) and $\alpha$ (azimuthal
angle) define the orientation in space of the line-of-sight (l.o.s.) 
towards observer. 
The label `Velocity' denotes the general direction of the shocked 
plasma flow.
The two dashed-line arrows illustrate that emission from a parcel of 
gas in the CSW region is subject to different wind absorption,
depending on the orientation ($i$, $\alpha$) of the l.o.s. towards 
observer  and the rotational angle around the axis of symmetry (Z).
{\it 
Right-hand panel:
}
the CSW region derived from the 2D hydrodynamic simulations for the
adopted values of the stellar-wind parameters in \WR which give a
ram-pressure ratio of the stellar winds 
$\Lambda = (\dot{M}_{WC8} V_{WC8})/(\dot{M}_{WN8} V_{WC8} = 1.44$.
The x- and z-axis are in units of the binary separation. The WN and
WC components are located at (x, z) coordinates (0,0) and (0,1),
respectively. The solid lines mark the shock fronts and the dashed
line marks the contact discontinuity.
}
\label{fig:cartoon}
\end{figure*}

\section{CSW Spectral modelling}
\label{sec:modelling}
As mentioned above, the main goal of this study is to carry out a
direct comparison of the X-ray spectrum of a massive binary with the
CSW model. For a more complete comparison, we focus mostly on the 
spectra with high spectral resolution, namely, on modelling the 
first-order \ChandraE-HETG spectra of \WRE, because they provide 
also pieces of kinematic information (through line profiles) of the 
X-ray emitting plasma.
In the global spectral modelling discussed here, we make use of 
the $\chi^2$ statistic and adopt the default standard weighting 
as defined in \xspecE.

\subsection{CSW model}
\label{sec:csw}
The basic feature of the standard CSW model in massive binaries is
that the two spherically symmetric stellar winds have collided at 
their terminal velocities.
Therefore, the numerical hydrodynamic model is 
two-dimensional (2D). Fig.~\ref{fig:cartoon} shows a schematic diagram 
of CSWs in a wide WC $+$ WN massive binary (i.e., \WRE, see
Section~\ref{sec:params}).

Our \xspec CSW model is based on the 2D numerical hydrodynamic model 
of adiabatic CSW by \citet{lm_90} (see also \citealt{mzh_93}). It can 
take into account partial electron heating in strong shocks (see
\citealt{zhsk_00}), non-equilibrium ionization (NEI) effects (see
\citealt{zh_07}), line broadening due to the bulk gas velocity of the
emitting plasma (see \citealt{zhp_10}), the specific stellar wind
absorption of both binary components along the line of sight to the
observer (see \citealt{zh_21}), and the different chemical composition
of both stellar winds.

For a detailed description of the CSW model and
the fitting procedure with the CSW model 
in \xspecE, we refer to section 4.1 of \citet{zh_17} and section 4.1 
of \citet{zh_21}.

\subsection{Adopted CSW model parameters for \WR}
\label{sec:params}
As known (e.g., \citealt{mzh_93} and references therein),
the mass-loss rate and velocity of the stellar winds of the binary 
components and the binary separation determine the shape and structure 
of the CSW region. So, these are the basic input parameters for the 
hydrodynamic CSW model in massive binaries.

We note that \citet{zh_etal_14} reported that \WR has an composite
optical spectrum, which could be represented
by a sum of two WR spectra (WC8 and WN8h).
This analysis also 
showed that there are two considerably different gas flows in this object. Namely,
the `cool' lines  (HeI and HI lines) had full width at half maximum of
FWHM $\approx 1000$\kms, while those of high-excitation ionic species
(e.g., CIV) have FWHM $\approx 2000$\kms (see fig.6 in
\citealt{zh_etal_14}).  So, \WR is considered a WC8 +
WN8 massive binary (e.g., see  \citealt{zh_etal_14}).

For the mass-loss rates of the binary components, we adopted the mean
values for the single WC8 and WN8 stars with known {\it Gaia} 
distances (see table 1 in \citealt{sander_19} and table 1 in 
\citealt{hamann_19}). 

To estimate the binary separation in \WRE, we adopted an orbital
period of $\approx 32$~yr \citep{williams_12} and rescaled the
semimajor axis of the prototype CSW binary WR 140 based on the
Kepler's third law and using the orbital parameters of WR 140 from
\citet{monnier_11}: orbital period of 2896 days (7.93 yr) and semimajor
axis of 14.73 au.

The adopted values of the stellar wind parameters and binary
separation in \WR are: 
$\dot{M}_{WC8} = 2.92\times10^{-5}$ and
$\dot{M}_{WN8} = 4.05\times10^{-5}$ (\dotMyr);
V$_{WC8} = 2000$ and V$_{WN8} = 1000$  (\kms);
semimajor axis of 37.3 au.

Based on these parameters, we see that the CSW shocks in \WR are
adiabatic (using equation 9 in \citealt{mzh_93}), the shock-heated
plasma may have different electron and ion temperatures (T$_e \neq$
T$_i$; using equation 1 in \citealt{zhsk_00}), and the NEI effects
are not important (using equation 1 in \citealt{zh_07}) that is the 
hot plasma in \WR is in collisional ionization equilibrium.

For consistency with the previous X-ray studies (\citealt{zhgsk_11};
\citealt{zhgsk_14}), we adopted a distance of 4 kpc to \WRE. We have 
to keep in mind that the {\it Gaia} DR2 (Data Release 2) distance to 
this object is not tightly constrained: 
$2.70 ^{+1.23}_{-0.67}$ kpc \citep{bailer_jones_18};
$2.27 ^{+0.92}_{-0.57}$ kpc \citep{rate_crowther_20}.
Also, we might expect some appreciable changes in the distance
estimates of \WR based on the parallax values of $0.3451\pm0.1082$ 
mas (DR2; \citealt{gaia_dr2}) and $0.1933\pm0.0462$ mas ({\it Gaia} 
EDR3, Early Data Release 3; \citealt{gaia_edr3}).

\begin{table}

\caption{\WR Spectral Fit Results (abundances)
\label{tab:abunds}}
\begin{center}
\begin{tabular}{lll}
\hline
\multicolumn{1}{c}{ } &
\multicolumn{1}{c}{$\beta = 1.0$ }  &
\multicolumn{1}{c}{$\beta = 0.2$ }\\
\hline
Mg &  0.47 (0.03) & 0.48 (0.03) \\
Si &  0.49 (0.01) & 0.48 (0.01) \\
S  &  1.06 (0.01) & 1.04 (0.01) \\
Ar &  0.28 (0.01) & 0.27 (0.01) \\
Ca &  0.16 (0.01) & 0.16 (0.02) \\
Fe &  0.44 (0.01) & 0.51 (0.01) \\
\hline

\end{tabular}
\end{center}

{\it Note}. 
Abundance values derived from the CSW model
simultaneous fits to the \ChandraE-MEG and HEG spectra.
Labels $\beta = 1.0$ and $\beta = 0.2$ denote
correspondingly the cases of full temperature equilibration and
partial electron heating at the shock fronts.
Given are the mean value for
each element and its standard deviation for the total number of 26
model fits (13 values of $\alpha \in [0, 180] $~degrees and 2 values
of $i = 60, 90$~degrees, 
the derived abundances are with respect to their typical values
adopted in this study, 
see Section~\ref{sec:fit_results}).

\end{table}

\begin{figure}
\begin{center}
\includegraphics[width=\columnwidth]{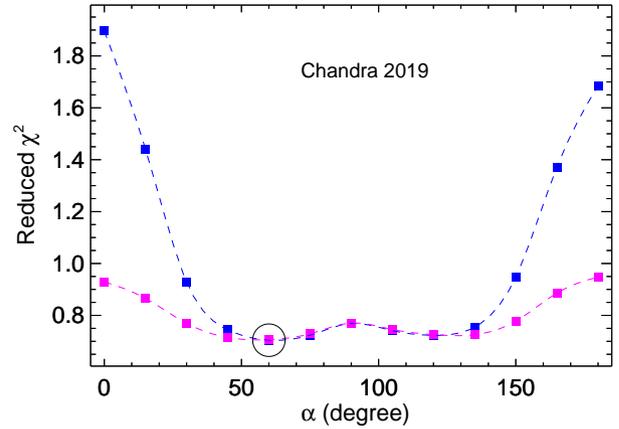}
\end{center}
\caption{
The reduced-$\chi^2$ values (degrees of freedom, dof = 251) vs.
azimuthal angle ($\alpha$, see Fig.~\ref{fig:cartoon}) for the case of
equal electron and ion temperatures ($\beta = 1$). The results for
inclination angle of $i = 90$ and $60$ degrees are shown in blue and
magenta colour, respectively. The black circle marks the formal
minimum value of the reduced $\chi^2$ at $\alpha = 60$ degrees.
}
\label{fig:chi2}
\end{figure}

\subsection{CSW model spectral results}
\label{sec:fit_results}
Since the orbital parameters of \WR are not known and we have only an
estimate of the binary separation (see Section~\ref{sec:params}), we
explored a range of values for the azimuthal angle and orbital
inclination (Fig.~\ref{fig:cartoon}). We considered 13 equidistantly 
spaced values of $\alpha \in [0, 180]$ degrees and two values of 
$i = 60, 90$ degrees. Due to the symmetry of the CSW region that 
resulted from interaction of two spherically-symmetric stellar winds, 
the CSW models with a given value of $\alpha$ and $360 - \alpha$ give 
identical spectra.
To see whether the partial electron heating at shock fronts could have
any impact on the X-ray emission from \WRE, we considered the cases of
$\beta = 1, 0.2$ ($\beta = T_e / T$, $T_e$ is the
electron temperature and $T$ is the mean plasma temperature).

We note that the X-ray spectrum of the CSW region is a sum of the
X-ray emission from both shocked winds, which each may have different
chemical composition.
And, we recall that the CSW model is capable of taking into account 
the different chemical composition in both parts of the interaction 
region.

For the shocked WC and WN wind in the CSW region of \WRE,
we correspondingly adopted the abundance values typical for the WC and WN 
stars (by number) as from \citet{vdh_86}.
Ar and Ca are not present in the \citet{vdh_86} abundance sets, so, we
adopted for each of them a fiducial value of $2\times10^{-5}$.

For the WC shocked wind we adopted:
H = 0.0, He = 1.0, C = 0.4, N = 0.0, O = 0.194,
Ne = $1.86\times10^{-2}$, Mg = $2.72\times10^{-3}$,
Si = $6.84\times10^{-4}$, S = $1.52\times10^{-4}$,
Ar = $2\times10^{-5}$, Ca = $2\times10^{-5}$,
Fe = $3.82\times10^{-4}$.

And, for the WN shocked wind we adopted:
H = 0.067, He = 1.0, C = $1.28\times10^{-4}$, 
N = $.29\times10^{-3}$, O = $2.92\times10^{-4}$, 
Ne = $6.57\times10^{-4}$, Mg = $2.19\times10^{-4}$, 
Si = $2.16\times10^{-4}$, S = $5.11\times10^{-5}$, 
Ar = $2\times10^{-5}$, Ca = $2\times10^{-5}$,
Fe = $1.28\times10^{-4}$.

It is worth mentioning that both shocked winds have about comparable
contribution to the total observed X-ray emission (flux) of \WR 
(see below). This is not surprising given the wind parameters
of both stellar components that result in comparable ram pressure (see
Section~\ref{sec:params} and Fig.~\ref{fig:cartoon}) and the fact that
both winds are fast (e.g., were one of the winds slow, say $\sim
100$\kms, its shocked plasma would not have been a strong X-ray 
source).
Also, for better quality of the fits, we allowed some abundances 
(Mg, Si, S, Ar, Ca, Fe) to vary.
Since the X-ray emission from the shocked WC and WN winds cannot be
disentangled, the abundance of a given element was varied by a single 
scaling parameter for both parts of the CSW region with respect to 
their reference abundances.
These are the derived abundance values from the spectral 
fits.

Because chemical abundances are best constrained from dispersed X-ray 
spectra, we fitted simultaneously the MEG and HEG spectra of \WRE.
The fitting procedure consisted of the following steps,
adopting our custom CSW model ($csw\_lines\_wind$) in \xspec that
takes into account line-broadening  due to the bulk gas velocity of
the hot plasma as well as stellar wind absorption
along the line of sight.

\begin{figure*}
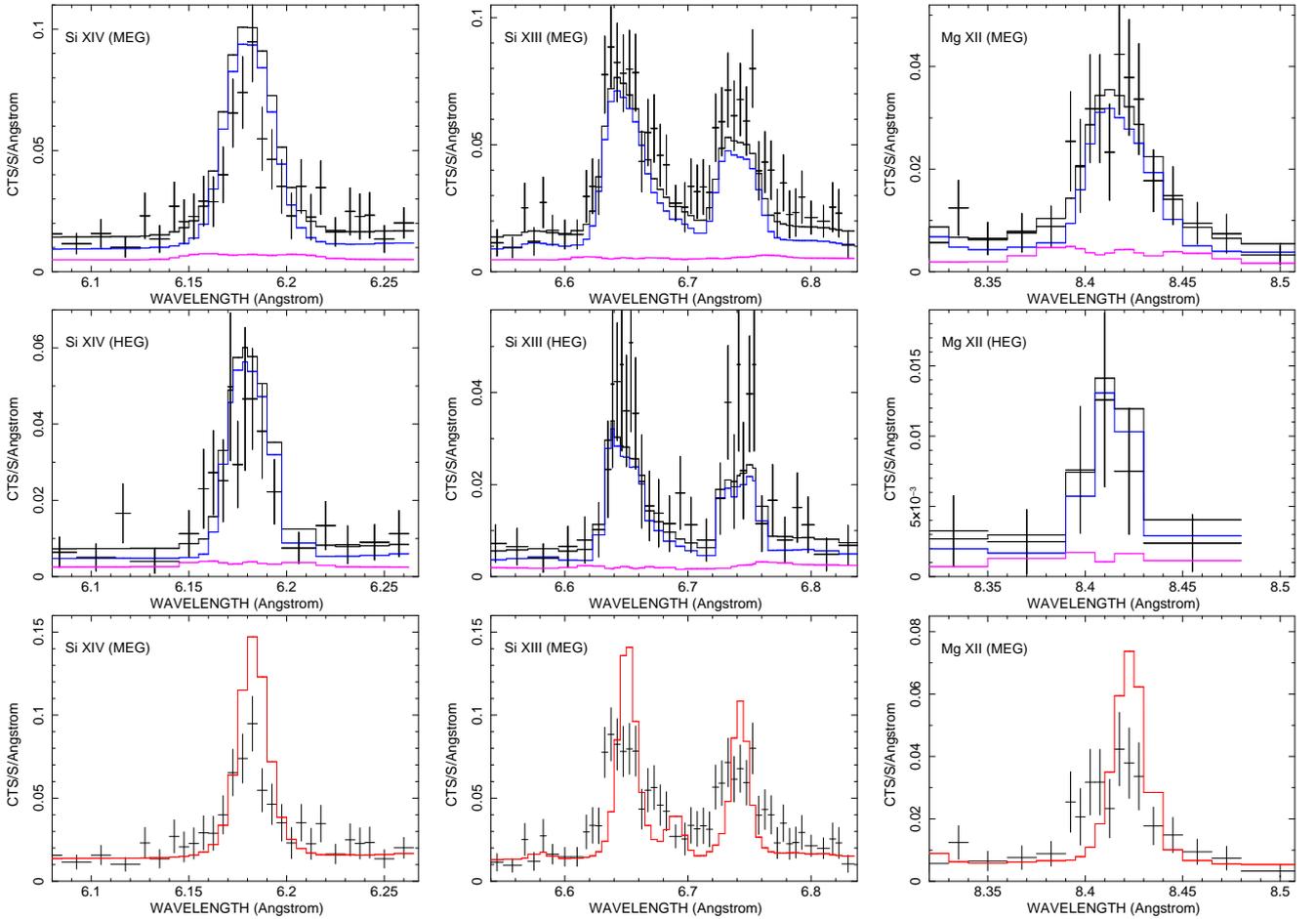

\begin{center}
\includegraphics[width=1.6in,height=2.3in,angle=-90]{fig6a.eps}
\includegraphics[width=1.6in,height=2.3in,angle=-90]{fig6b.eps}
\includegraphics[width=1.6in,height=2.3in,angle=-90]{fig6c.eps}
\includegraphics[width=1.6in,height=2.3in,angle=-90]{fig6d.eps}
\includegraphics[width=1.6in,height=2.3in,angle=-90]{fig6e.eps}
\includegraphics[width=1.6in,height=2.3in,angle=-90]{fig6f.eps}
\includegraphics[width=1.6in,height=2.3in,angle=-90]{fig6g.eps}
\includegraphics[width=1.6in,height=2.3in,angle=-90]{fig6h.eps}
\includegraphics[width=1.6in,height=2.3in,angle=-90]{fig6i.eps}
\end{center}
\caption{
{\it Top and middle panels.} 
The MEG and HEG background-subtracted spectra of \WR and the
best-fit CSW model ($\alpha = 60$ degrees) near some strong emission lines. 
The WN and WC shocked stellar wind spectra are plotted in
blue and magenta colour, respectively.
{\it Bottom panels.} For comparison, shown are the MEG spectra 
overlaid with the CSW model for azimuthal angle $\alpha = 0$ degrees
and inclination angle $i = 90$ degrees (the case of reduced $\chi^2 =
1.9$, see Fig.~\ref{fig:chi2}).
The spectra were rebinned to have a minimum of 10 counts per bin.
}
\label{fig:lines_csw}
\end{figure*}

(1) For each individual case ($\alpha, i, \beta$), 
we fitted the high-resolution spectra to estimate the
abundances (the MEG and HEG spectra were rebinned to have a minimum 
of 20 counts per bin).
To minimize the amount of CPU time, 
we used the \xspec model {\it gsmooth} for the line broadening
with a constant FWHM $= 1400$\kms over the entire spectrum 
(see Section~\ref{sec:lines} and Fig.~\ref{fig:lines_param}).
So, the fitted \xspec model is:
$Spec = tbabs * gsmooth(csw\_lines\_wind)$
as the line-broadening was switched off in the $csw\_lines\_wind$ 
model. 
The {\it tbabs} model accounts for the interstellar 
(and circumstellar) absorption.
The fit results showed that the derived abundance values for each
chemical element have no big scatter: they have relatively small 
standard deviation around their corresponding mean value
(see Table~\ref{tab:abunds}).

(2) 
For each value of $\beta$, we used the corresponding mean
abundance values from Table~\ref{tab:abunds} and we repeated all the
spectral fits as described in step (1) with abundances kept fixed.
We thus
derived typical values for other CSW parameters: e.g., mass-loss
scaling factor, X-rays absorption. It is worth noting that for both
cases of $\beta$ the fractional mass-loss reduction was 
$\dot{M}_s \approx 0.27$ with respect to the nominal mass-loss 
values adopted in this study (see Section~\ref{sec:params}).
On the other hand, the X-ray absorption is different between the cases
of $\beta = 1$ and $0.2$ and it also varies with the azimuthal angle.
All this is well understood and we will further discuss it in
Section~\ref{sec:discussion}.

(3) The best-fit models from step (2) were then used to check whether
the CSW model provides the right kinematics of the X-ray emitting
plasma in \WRE. Namely, 
the fitted \xspec model is:
$Spec = tbabs * csw\_lines\_wind$ as the line broadening in the CSW
model was switched on.
Since the spectral line profiles provide information on
the gas kinematics of the X-ray emitting plasma,
we used the same spectral ranges for 
various emission lines as in the standard line fitting (see 
Section~\ref{sec:lines}) to estimate correspondence between theory and
observations, and these lines were considered (analysed) not
one-by-one but simultaneously. 
Also, as a trade-off between spectral resolution and
spectral quality we applied these models to the MEG and HEG spectra
rebinned to have a minimum of 10 counts per bin.

Figure~\ref{fig:chi2} presents the $\chi^2$ values for
all the 26 cases under consideration 
and equal electron and ion temperatures ($\beta = 1$).
We note that we found no appreciable difference in the corresponding
$\chi^2$ values between the cases of $\beta = 1$ and $\beta = 0.2$: 
the differences were less than 1\%. Although the formal minimum of the
reduced $\chi^2$ is at azimuthal angle of $\alpha = 60$ degrees, 
we see that models in a broad range $\alpha \in [ 45, 135]$ degrees 
could be considered acceptable for the quality of the 2019 
\ChandraE-HETG spectra of \WRE.

Examples of the direct confrontation of the observed line profiles in
the X-ray spectrum of \WR and the CSW model are shown in
Fig.~\ref{fig:lines_csw}. 
We see that although the CSW model does not provide a perfect match to 
the observed line profiles, the theoretical line profiles could be 
considered an acceptable representation of the
observed line profiles of the strong line features in the X-ray
spectrum of this CSW binary.
An illustration of not acceptable theoretical profiles is shown in the
bottom panels of Fig.~\ref{fig:lines_csw} (the case of azimuthal
angle $\alpha = 0$ degrees and inclination angle  $i = 90$ degrees;
the case of reduced $\chi^2 = 1.9$ in Fig.~\ref{fig:chi2}). These 
profiles are narrower than the observed ones and slightly redshifted 
with respect to them.
This is a result from the considerably large opening angle of the CSW
cone (see right-hand panel in Fig.~\ref{fig:cartoon}). We have to keep
in mind that the CSW cone has axial symmetry, but it is a 3D object. 
So, a parcel of hot gas, having the same plasma parameters, might be 
subject to different wind absorption depending on the rotational angle 
around the axis of symmetry of the CSW cone and the line-of-sight 
towards observer (e.g., left-hand panel in Fig.~\ref{fig:cartoon}). 
This may result in symmetric or asymmetric line profiles and all these 
details are taken into account by the CSW model.

An interesting feature of the CSW model is that it can provide the
contribution of each shocked stellar wind to the total X-ray emission
from a CSW binary. For the case of \WRE, we find that both shocked
stellar winds have similar contribution to the total observed X-ray
flux in the (0.5 - 10 keV) energy range: 54\% (WN) and 46\% (WC).
But, the emission of the strong line features is dominated by the 
shocked WN stellar wind as illustrated in Fig.~\ref{fig:lines_csw}.
This could be understood in the way that the plasma temperature in the
WC part of the CSW region is considerably higher than that in the WN
shocked plasma, thus, the WC shocked gas contributes mostly to the
continuum emission. On the other hand, the WN shocked wind has the
`right' temperature to boost line emission. Of course, this is the
case for the particular values of the wind velocities and chemical
compositions adopted in this sudy.

\begin{table*}

\caption{\WR Spectral Fit Results 2019 - 2008
\label{tab:fits}}
\begin{center}
\begin{tabular}{lcccccc}
\hline
\multicolumn{1}{c}{ } &
\multicolumn{3}{c}{$\beta = 1.0$ }  &
\multicolumn{3}{c}{$\beta = 0.2$ }\\
\multicolumn{1}{c}{ } &
\multicolumn{1}{c}{2019 }  &
\multicolumn{1}{c}{2012 }  &
\multicolumn{1}{c}{2008 }  &
\multicolumn{1}{c}{2019 }  &
\multicolumn{1}{c}{2012 }  &
\multicolumn{1}{c}{2008 }  \\

\hline
$\chi^2$ (min - max)  & 274 - 288 & 56 - 57 & 809 - 994 & 
            265 - 283 & 54 - 55 & 645 - 778 \\
\hspace{0.3in} dof & 567 & 79 & 611 & 567 & 79 & 611 \\
$\dot{M}_s$ (mass-loss reduction) &
          0.27 (0.01) & 0.20 (0.01) & 0.28 (0.01) &
          0.27 (0.01) & 0.21 (0.01) & 0.29 (0.01) \\
$F_X$ ($10^{-11}$ ergs cm$^{-2}$ s$^{-1}$)  &
          0.95 (0.01) &  0.36 (0.01) &  1.01 (0.01) &
          0.91 (0.01) &  0.34 (0.01) &  0.98 (0.01) \\
$F_0$ \,\,($10^{-11}$ ergs cm$^{-2}$ s$^{-1}$) &
          4.53 (0.11) &  2.60 (0.04) &  5.02 (0.16) &
          5.00 (0.12) &  2.94 (0.04) &  5.64 (0.17)  \\
$\log L_X$ (ergs s$^{-1}$) &
          34.94 & 34.70 & 34.98 & 34.98 & 34.75 & 35.03 \\
\hline

\end{tabular}
\end{center}

{\it Note}. Results from the CSW model fits to the X-ray spectra of 
\WRE:: Chandra 2019 (columns marked with 2019); Chandra 2012
(columns marked with 2012); XMM-Newton 2008 (columns marked with
2008). Labels $\beta = 1.0$ and $\beta = 0.2$ denote 
correspondingly the cases of full temperature equilibration and partial 
electron heating at the shock fronts. 
Tabulated quantities are the mass-loss scaling factor ($\dot{M}_s$),
the observed X-ray flux ($F_X$), the unabsorbed (net) X-ray flux 
($F_0$) and the logarithm of the X-ray luminosity ($\log L_X$) for an 
adopted distance of 4 kpc to \WRE. The last three quantities are in 
the 0.5 - 10 keV energy range.
Given are the mean values and their standard deviation for the total 
number of 26 model fits (13 values of $\alpha \in [0, 180] $~degrees 
and 2 values of $i = 60, 90$~degrees, see Section~\ref{sec:fit_results}).

\end{table*}

\section{Discussion}
\label{sec:discussion}
We carried out a direct modelling of the observed X-ray spectra
with high spectral resolution (\ChandraE-HETG) of \WR in the 
framework of the standard colliding stellar wind picture 
in a massive WC+WN binary. 
Two of the most interesting CSW model results are the following.

First, a fractional mass-loss reduction of $\dot{M}_s \approx 0.27$
with respect to the nominal mass-loss values adopted in this
study (see Section~\ref{sec:params}) is needed to match the observed
X-ray flux. It was a result from the fact that the CSW model with the
nominal mass-loss values predicted too high an emission measure (EM) 
for the X-ray plasma. 
It is worth recalling that in the case of spherically-symmetric
stellar winds the emission measure in the CSW region 
is proportional to the square of the stellar wind mass loss ($\dot{M}$) 
and is reversely proportional to the binary separation ($a$):
EM $\propto \dot{M}^2/a$ (e.g., see section 4.2 in \citealt{zh_17}).
We note that this mass-loss reduction is related to the adopted 
distance of $d = 4$ kpc to \WR (see Section~\ref{sec:params}). 
So, if the actual distance to \WR is smaller or larger than 4 kpc, 
then the amount of emission measure needed to match the observed X-ray 
flux will correspondingly decrease or increase ($\propto d^2$), 
so will the mass-loss reduction with respect to the value of 
$\dot{M}_s \approx 0.27$ ($\propto d$).

Second, the azimuthal angle of the observer was $\alpha = 60$ degrees 
(see Fig.~\ref{fig:cartoon}) or more likely it was in the range 
$\alpha \in [45, 135]$ degrees as of 2019 November-December. On the
other hand, 
the current knowledge of the stellar wind and binary parameters \WR 
did not allow us to obtain any constraints on the inclination angle 
of its binary orbit.

\begin{figure}
\begin{center}
\includegraphics[width=\columnwidth]{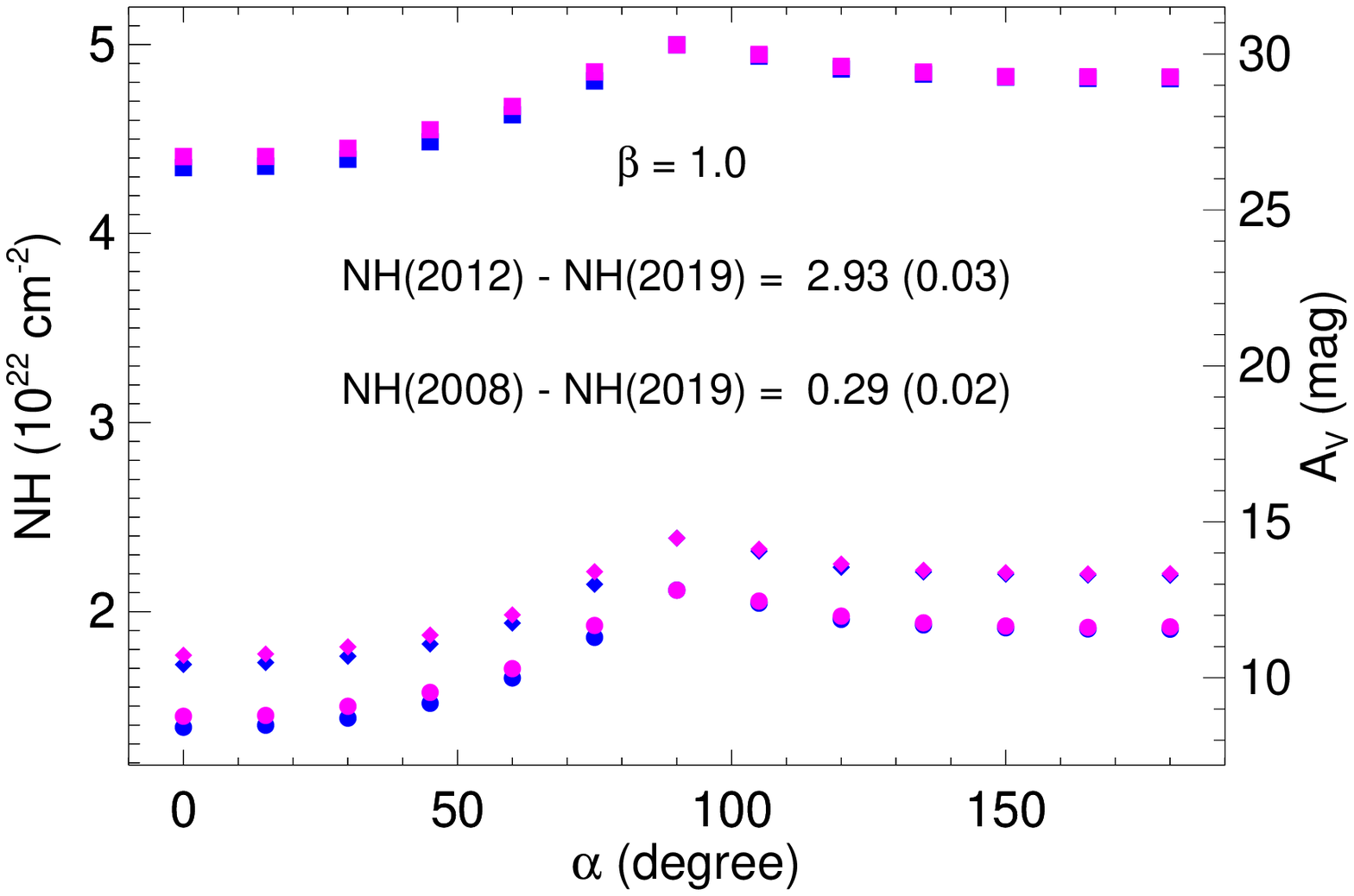}
\includegraphics[width=\columnwidth]{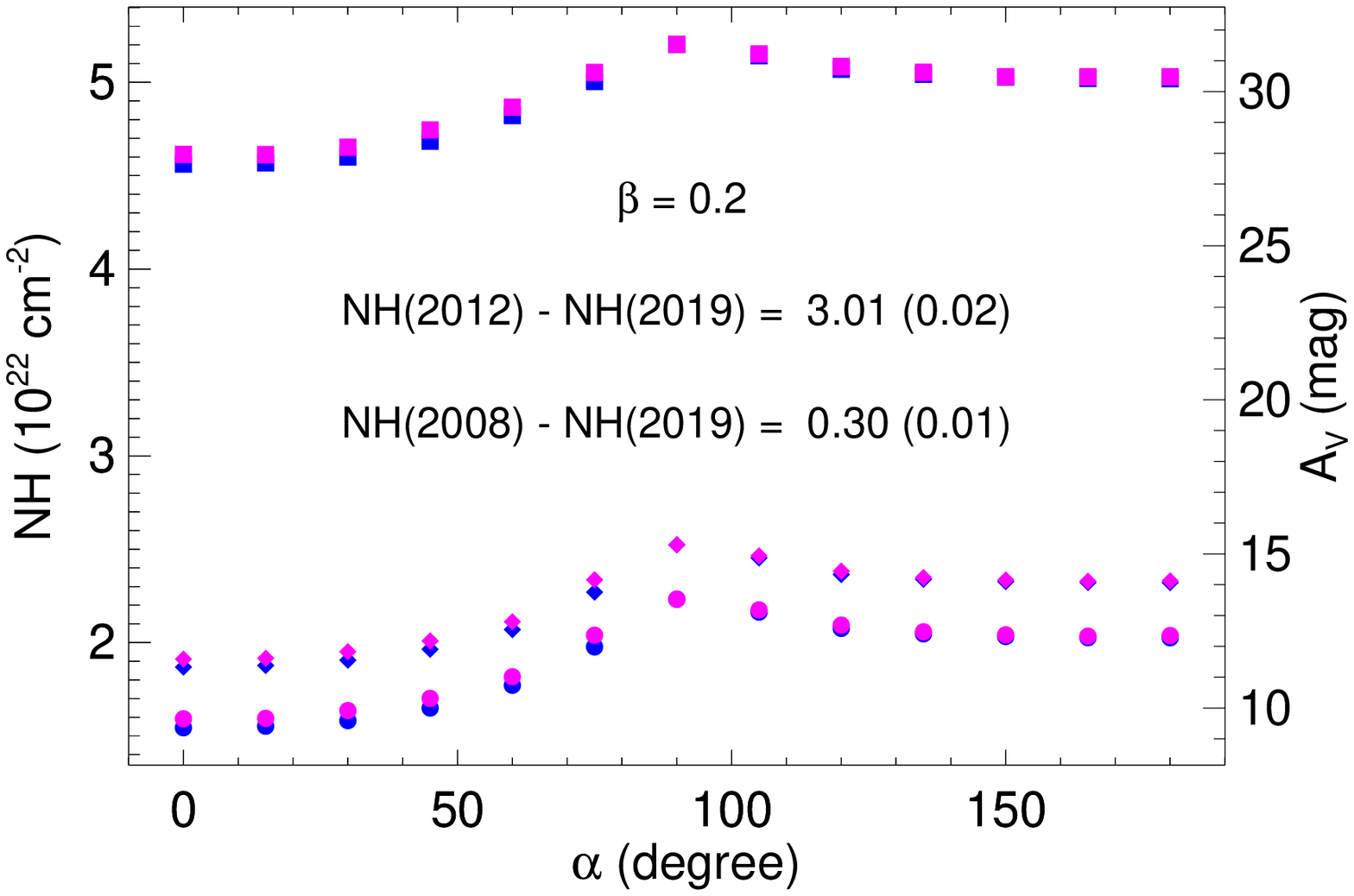}
\end{center}
\caption{
The X-ray absorption (NH) vs. azimuthal angle ($\alpha$, see
Fig.~\ref{fig:cartoon}).
The results for inclination angle of $i = 90$ and $60$ degrees
are shown in blue and magenta colour, respectively.
The bottom curves are for  Chandra 2019, the curves in the middle are
for XMM-Newton 2008 and the top curves are for Chandra 2012. The mean
differences of the X-ray absorption and their standard deviations (in
parentheses) are given as well.
}
\label{fig:nh}
\end{figure}

However, in order to derive more constraints on the general CSW 
picture in \WR we think it could be a good idea to  expand our current 
CSW-model analysis to other observations with good (or acceptable) 
quality. To do so, we made use of the previous X-ray observations of 
\WR as of 2008 January (\XMME) and 2012 October (\ChandraE). Details 
on the standard X-ray analysis of these data sets are found in
\citet{zhgsk_11} and \citet{zhgsk_14}. We may refer to these data sets
as `XMM-Newton 2008' and `Chandra 2012' throughout the text.

Chandra 2012. This is a \ChandraE-HETG observation of \WR carried out 
on 2012 October 12 (\Chandra Obs ID 13636). For the purpose of this
analysis, we re-extracted the MEG and HEG spectra following the same
data reduction recipe as for the Chandra 2019 data set (see
Section~\ref{sec:data}).
In the CSW model analysis, we made use of MEG and HEG spectra both
re-binned to have a minimum of 10 and 20 counts per bin
and we followed step (2) and (3) as described in 
Section~\ref{sec:fit_results} (i.e., adopting the abundance sets from
Table~\ref{tab:abunds}). Due to the quality of the data (see
Section~\ref{sec:lines}; also fig. 2 in \citealt{zhgsk_14}), 
only the MEG spectrum near the S XV, Si XIV and Si XIII lines was
used in the step (3) of this analysis.

XMM-Newton 2008. This is an \XMM observation of \WR carried out on 
2008 January 9 (\XMM Obs ID 0510980101). For the purpose of this 
analysis, we made use of the data from the pn detector of the European 
Photon Imaging Camera (EPIC). We used the \XMM Science Analysis System 
(SAS, version 16.1.0)\footnote{The \XMM Science Analysis System (SAS),
\url{https://www.cosmos.esa.int/web/xmm-newton/sas}.} to filter the
data for high X-ray background and to re-extract the source and 
background spectra and their corresponding response files.
In the CSW model analysis, we used the pn spectrum 
re-binned to have a minimum of 100 counts per bin
and we followed step (2) as described in
Section~\ref{sec:fit_results}, also adopting the abundance sets from
Table~\ref{tab:abunds}.

\begin{figure}
\begin{center}
\includegraphics[width=\columnwidth]{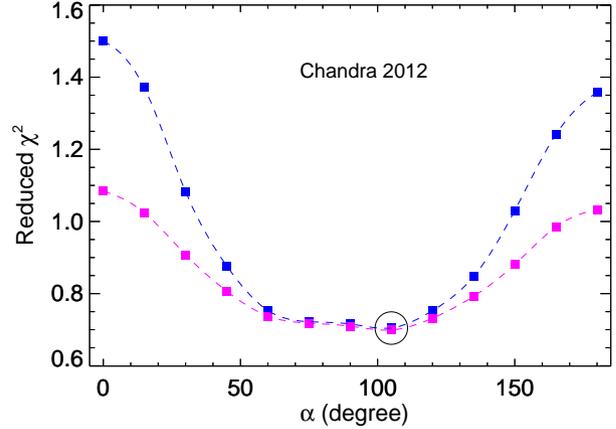}
\end{center}
\caption{
The reduced-$\chi^2$ values (degrees of freedom, dof = 31) vs.
azimuthal angle ($\alpha$, see Fig.~\ref{fig:cartoon}) for the case of
equal electron and ion temperatures ($\beta = 1$). The results for
inclination angle of $i = 90$ and $60$ degrees are shown in blue and
magenta colour, respectively. The black circle marks the formal
minimum value of the reduced $\chi^2$ at $\alpha = 105$ degrees.
}
\label{fig:chi2_2012}
\end{figure}

\begin{figure*}
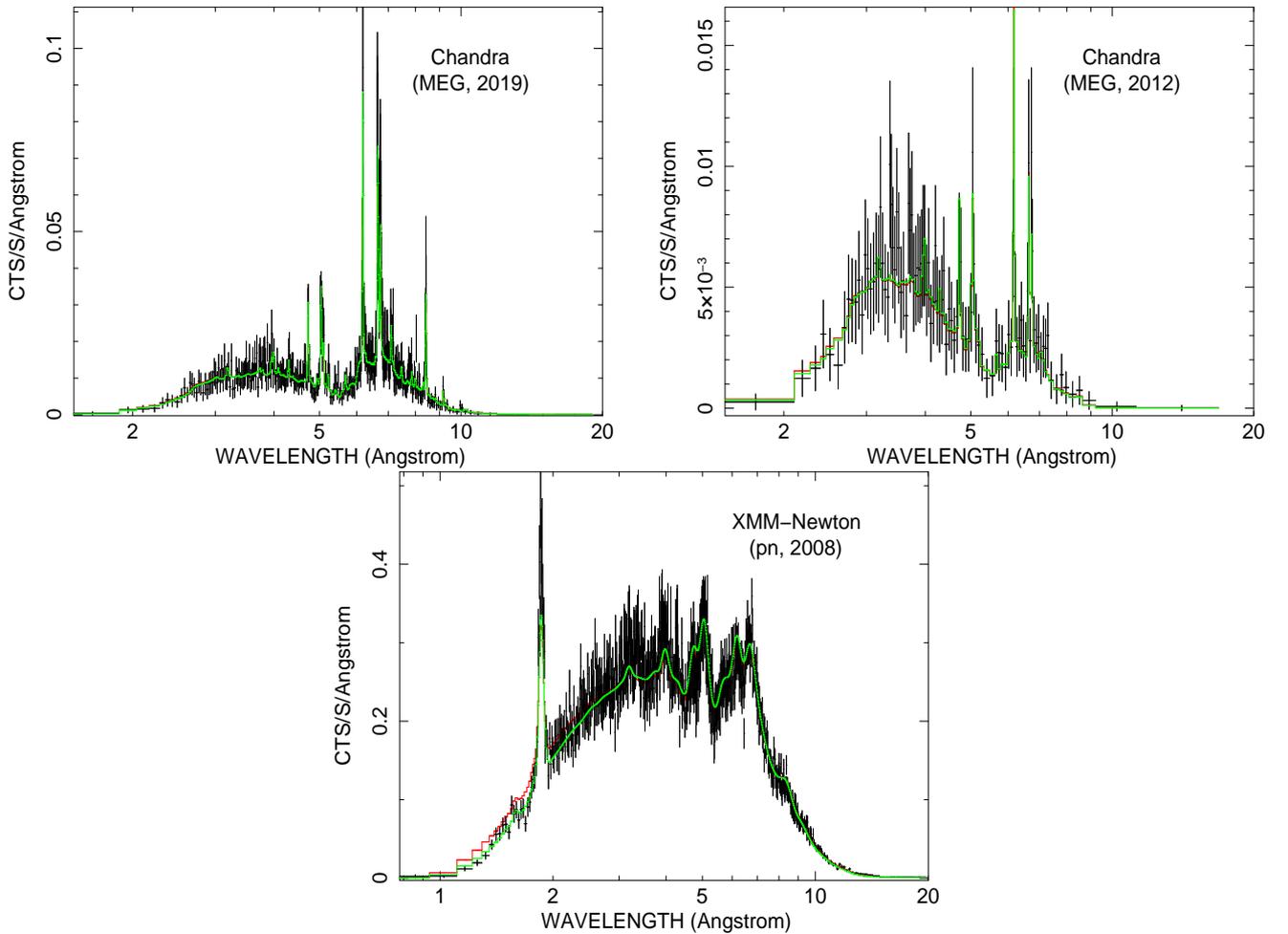

\begin{center}
\includegraphics[width=2.5in,height=3.5in,angle=-90]{fig9a.eps}
\includegraphics[width=2.5in,height=3.5in,angle=-90]{fig9b.eps}
\includegraphics[width=2.5in,height=3.5in,angle=-90]{fig9c.eps}
\end{center}
\caption{The background-subtracted spectra of \WR from different
epochs overlaid with the best-fit CSW model. The \Chandra  and \XMM
spectra were re-binned to have a minimum of 20 and 100 counts per bin,
respectively.
The model spectra for $\beta = 1$ and $0.2$ are plotted with red and
green colour, respectively.
}
\label{fig:spectra}
\end{figure*}

Some results from the CSW-model fits to the Chandra 2012 and
XMM-Newton 2008 spectra along with those for the Chandra 2019 spectra
are given in Table~\ref{tab:fits},
Figs.~\ref{fig:nh},~\ref{fig:chi2_2012}  and \ref{fig:spectra}. 
We immediately notice that the X-ray luminosity of \WR derived from 
the CSW-model analysis (Table~\ref{tab:fits}) confirms its status of
the most X-ray luminous Wolf-Rayet star in the Galaxy detected so far,
after the black-hole candidate Cyg X-3 \citep{zhgsk_11}.

In the framework of the CSW picture, we also see that the observed
characteristics of the X-ray emission region (i.e., the CSW zone) in 
\WR are quite similar in 2019 and 2008. On the other hand, it is 
confirmed what was already reported \citep{zhgsk_14} that the X-ray 
emission from \WR was appreciably lower in 2012 compared to its level 
in 2008. The same is valid for the amount of emission measure (EM 
$\propto \dot{M}_s^2$). It is also confirmed that the decrease of the 
emission measure in 2012 was accompanied with a considerable increase 
of the X-ray absorption. 

We emphasize that the derived amount of X-ray 
absorption (Fig.~\ref{fig:nh}) is in addition to that due to the 
stellar winds, that is it is of interstellar and circumstellar origin. 
The optical extinction toward \WR has been reported to be very high, 
A$_V = 9.2$~mag (\citealt{danks_83}; see also fig. 8 in
\citealt{zh_etal_14}), and we note that a conversion N$_H =
1.65\, (1.6-1.7)\times10^{21}$A$_V$~cm$^{-2}$ (\citealt{vuong_03}; 
\citealt{getman_05}) is used in Fig.~\ref{fig:nh} 
(for the right-hand y-axis).

The apparent variation of the excess X-ray absorption with the
azimuthal angle is a result from different chemical composition of the
stellar winds. Namely, the more chemically evolved WC wind has a
higher X-ray absorption than that of the WN wind, so, the additional
X-ray absorption needed to match the shape of the X-ray spectrum is
therefore smaller for azimuthal angles of $\alpha < 90$~degrees. The
opposite is valid for azimuthal angles of $\alpha > 90$~degrees, when
the WN component is predominantly `in front'. Naturally, the peak of
the excess X-ray absorption is near $\alpha \approx 90$~degrees, when
the stellar wind absorption is minimal (the observer's line-of-sight is
approximately `through' the CSW region itself, i.e. along the x-axis
in Fig.~\ref{fig:cartoon}).

Figure~\ref{fig:chi2_2012} shows the results from confronting the
CSW model profiles with the observed ones for Chandra 2012 for the
case of equal electron and ion temperatures ($\beta = 1$). As in the
case of Chandra 2019 (see Section~\ref{sec:fit_results} and
Fig.~\ref{fig:chi2}), we found no appreciable difference in the
corresponding $\chi^2$ values  between the cases of $\beta = 1$ and
$\beta = 0.2$: the differences were less than 2\%. The 
formal minimum of the reduced $\chi^2$ is at azimuthal angle of
$\alpha = 105$ degrees, 
but similarly to the case of the 2019 \ChandraE-HETG spectra of \WRE, 
the models in a broad range 
(this time $\alpha \in [ 60, 120]$ degrees) could be considered acceptable.

Finally, the case of partial electron heating at
the strong shocks in the CSW region of \WR finds some support but {\it
only} from the case of XMM-Newton 2008. We see that the quality of the
CSW model fits for $\beta = 0.2$ is better than that for $\beta = 1$:
the $\chi^2$ values of the former are by $\sim 25$\% lower than those
for the latter. Also as seen from Fig.~\ref{fig:spectra}, while there
is practically no difference between the CSW model spectra with $\beta
= 1$ and $\beta = 0.2$ for Chandra 2019 and Chandra 2012, that with
equal electron and ion temperatures ($\beta = 1$) predicts hard X-ray 
emission (wavelengths $< 2 $\AA) higher than the observed one in 
XMM-Newton 2008.
Since partial electron heating at shock fronts ($\beta < 1$) results 
in 'on average' lower electron temperature in the CSW region, the much
larger effective area of the \XMME-EPIC instruments helps detect some
differences between X-ray spectra in the cases $\beta = 1$ and $\beta
= 0.2$.

In general, we see that the CSW-model analysis of the X-ray spectra 
(with good quality: high spectral resolution, high photon statistics) 
of \WR  provided some basic features of the CSW picture in this
WC$+$WN massive binary. 
Namely, the observed X-ray emission from \WR is
variable on long timescale (years) and the same is valid for its 
intrinsic X-ray emission. This requires variable mass-loss rates over 
its orbital period. And, the X-ray absorption is variable as well, 
as this absorption is in excess to that due to the stellar winds 
in the binary. It is worth noting that lower intrinsic X-ray 
emission is accompanied by higher X-ray absorption. Such a behaviour 
is quite similar to that of WR 140, which is the prototype of the 
CSW binaries showing periodic dust formation.

Namely, it is currently assumed that the dust formation in CSW
binaries is set up near the periastron passage as illustrated by the 
case of WR 140 (\citealt{williams_90}; also see \citealt{williams_95},
\citealt{williams_08}). Interestingly, the X-ray emission from this
object decreases at about the same orbital phases correlating with an
increase of the X-ray absorption (see \citealt{po_05};
\citealt{po_12}). 

An in-detail CSW-model analysis of the X-ray emission from WR 140 is 
presented in \citet{zh_21} showing that the mass-loss rates must 
decrease near periastron, and the following explanation was suggested.
`The variable effective mass-loss rate could be understood {\it
qualitatively} in CSW picture of clumpy stellar winds where clumps are
efficiently dissolved in the CSW region near apastron but not at
periastron. The increased X-ray absorption near periastron might 
be a sign of non-spherically symmetric stellar winds.'

In the framework of this qualitative CSW picture and following the
time sequence of the \WR X-ray observations, we may assume that \WR 
was observed near periastron in 2012 (Chandra 2012), while it was
observed quite a bit before and after periastron in 2008
(XMM-Newton 2008) and 2019 (Chandra 2019), respectively. The presumed
periastron passage was likely at the end of 2011 (beginning of 2012)
as maximum in the \WR light curve in the near infrared indicates 
(see fig. 3 in \citealt{williams_12}).
Interestingly, a minimum is present in the X-ray light curve of \WR at
about the same period of time (see fig. 10 in \citealt{zh_etal_14};
these short-exposure  observations were carried out with the \Swift 
observatory).
And, we could add another detail in this qualitative picture.

Our analysis of the X-ray data on \WR  with high spectral
resolution (Chandra 2019; Chandra 2012) showed that the azimuthal
angle of the line-ot-sight towards observer (Fig.~\ref{fig:cartoon}) 
was in the broad range of 
$\alpha \in [45, 135]$ and  $\alpha \in [60, 120]$ 
degrees in 2019 and in 2012, respectively.. Since such ranges are 
somehow `centred' at the value of 
$\alpha = 90$ degrees, it might be considered as an indication that 
this massive binary (\WRE) is observed `pole-on'. We note that 
due to the axial symmetry of the CSW region all the spectra (line
profiles)  for azimuthal angle of $\alpha = 90$ degrees and 
inclination angle $i \neq 0$ degrees are the same, and they are also
{\it identical} to those with inclination angle $i = 0$ degrees and 
arbitrary value of azimuthal angle. 
We have to keep in mind that even if \WR is observed `pole-on', it
will show variable X-ray emission provided its orbit has high
eccentricity.
By analogy with WR 140, the
variable X-ray absorption (being the highest near periastron) could be
a sign of non-spherically symmetric stellar winds. 

We believe that more X-ray observations of massive binaries with high 
spectral resolution and good photon statistics as well as future 
development of CSW models with non-spherically symmetric winds may 
help us get a deeper insight of the CSW picture in these objects.

\section{Conclusions}
\label{sec:conclusions}
The basic results and conclusions from our analysis of the X-ray
spectra of \WR with good quality (high spectral resolution, high
photon statistics) are as follows.

(i) Analysis of the line profiles of strong emission features in 
the X-ray spectrum of \WR from recent \ChandraE-HETG observations
(2019 November - December) showed that the spectral lines in this
massive binary are broadened (typical FWHM of 1400\kms) and 
marginally blueshifted by $\sim 100$\kms.

(ii) A direct modelling of these \Chandra (MEG, HEG) spectra in 
the framework of the standard CSW picture provided a very good
correspondence between the shape of the theoretical and observed
spectra. Also, it showed that the theoretical line profiles are in
most cases an acceptable representation of the observed ones. However, 
no tight constraints are derived on the azimuthal angle of the 
line-of-sight towards observer: it was in the range [45, 135] degrees 
at the time of observations.

(iii) To broaden this analysis, we applied the CSW model to the
X-ray spectra of \WR from previous observations: \ChandraE-HETG (2012
October) and \XMM (2008 January). The basic findings from the CSW
modelling of all the three data sets are the following.
The observed X-ray emission from \WR is variable on the long timescale 
(years) and the same is valid for its intrinsic X-ray emission. This 
requires variable mass-loss rates over the binary orbital period. The 
X-ray absorption is variable as well, as this absorption is in excess 
of that due to the stellar winds in the binary. Interestingly, lower 
intrinsic X-ray emission is accompanied by higher X-ray 
absorption.

(iv) The basic features described in the previous item are very
similar to those found in the prototype CSW binary WR 140
based on the CSW modelling (see \citealt{zh_21}). 
By analogy with WR 140, we propose the same qualitative CSW picture
for their explanation as given in \citet{zh_21}.
`The variable effective mass-loss rate 
could be understood in CSW picture of clumpy stellar winds where 
clumps are efficiently dissolved in the CSW region near apastron but 
not at periastron. The increased X-ray absorption near periastron 
might be a sign of non-spherically symmetric stellar winds.'

\section*{Acknowledgements}
This research has made use of data and/or software provided by the
High Energy Astrophysics Science Archive Research Center (HEASARC),
which is a service of the Astrophysics Science Division at NASA/GSFC
and the High Energy Astrophysics Division of the Smithsonian
Astrophysical Observatory. 
This research has made use of the NASA's Astrophysics Data System, and
the SIMBAD astronomical data base, operated by CDS at Strasbourg,
France.

Support for this work was provided by the National Aeronautics and
Space Administration (NASA) through Chandra Award Numbers GO9-20007A
(SLS), GO9-20007B (MG) and GO0-21015E (MG) issued by the Chandra X-ray
Center, which is operated by the Smithsonian Astrophysical Observatory
for and on behalf of NASA under contract NAS8-03060.
SAZ acknowledges financial support from
Bulgarian National Science Fund grant DH 08 12.
The authors thank the reviewer Dr Maurice A. Leutenegger for his
valuable comments and suggestions.

\section*{Data Availability}

The X-ray data underlying this research are {\it
public} and can be accessed as follows.
The \Chandra data sets can be downloaded from the \Chandra X-ray
observatory data archive
\url{https://cxc.harvard.edu/cda/} by typing in the target name (\WRE)
in the general search form \url{https://cda.harvard.edu/chaser/}.
The \XMM data sets can be downloaded from the \XMM Science Archive 
by typing in the
object name (\WRE) in the general search form 
\url{http://nxsa.esac.esa.int/nxsa-web/#search}.
%



\bibliographystyle{mnras}
\bibliography{wr48a} 


\bsp	
\label{lastpage}
\end{document}